# A Unified Device-Level Physical Framework for DRAM Read Disturbance

Zhaoqiang Bai[1], Haocong Luo[2], Xinhe Wang[1], Xujing Li[1], Yuelin Jiang[1],
Xuelu Huang[1], Guilei Wang[1], Hao Meng[1], Ataberk Olgun[2], Ismail Emir Yuksel[2],
F. Nisa Bostancı[2], Jikun Wang[2], and Onur Mutlu[2]

[1] Beijing Superstring Academy of Memory Technology, Beijing 100176, China
[2] ETH Zurich, 8092, Zürich, Switzerland

## Abstract

Read disturbance in dynamic random-access memory (DRAM) depends on both local device interactions and the electrical conditions imposed by shared array circuitry. Here, we combine DIMM-level characterization results, device measurements, and TCAD simulations to interpret RowHammer, RowPress, and ColumnDisturb within a unified device-level physical framework. We relate charge evolution at the floating storage node to three key mechanisms: trap-assisted electron migration (EM), capacitive coupling (CC), and sense-amplifier-mediated disturb (SAMD). EM is local and trap-sensitive, while CC weakens with wordline separation. SAMD extends disturbance through shared column circuitry by modifying unselected-cell leakage under rail-level bitline biases, a contribution supported by the approximately ninefold higher OFF-state leakage measured relative to retention-like conditions for a high-voltage storage state. These distinct spatial ranges account for localized RowHammer and RowPress failures and remote ColumnDisturb populations with abrupt changes at subarray boundaries. We further interpret prolonged RowPress activation as a transition from an initial RowHammer-like transient to sustained leakage driven mainly by SAMD under held wordline and bitline biases. Within this device-level framework, trap effectiveness and storage-state-dependent leakage provide an interpretation of previously puzzling failure signatures, including the earlier onset of $0 \rightarrow 1$ RowHammer failures despite the predominance of $1 \rightarrow 0$ failures at high hammer counts. The same framework explains the observed RowPress and ColumnDisturb polarities and the duration-dependent reversal of neighboring- and passing-wordline effectiveness. These results provide a physical basis for developing compact DRAM reliability models across access patterns and array configurations.

## 1. Introduction

Dynamic random-access memory assumes that reading one row preserves the information stored in all unselected rows. RowHammer breaks this isolation: repeated activation and precharge of an aggressor row can induce bitflips in nearby victim rows whose wordlines are never selected [1]. As DRAM has scaled, experimentally observed failure thresholds have decreased substantially [2], turning read disturbance from an anomalous device effect into a growing reliability and security constraint.

RowHammer is now only one manifestation of a broader disturbance problem. RowPress produces failures by keeping an aggressor row active for an extended interval rather than repeatedly switching it [3]. ColumnDisturb extends the affected region still further: activity in one row can disturb remote cells that share column circuitry across multiple subarrays [4]. Together with strong dependencies on data pattern, temperature, active time and cell location [1,5,6], these observations show that read disturbance cannot be described by activation count alone.

A common physical explanation must first account for where failures occur. The measurements examined here reveal a spatial hierarchy: RowHammer and RowPress produce strong failure peaks near the aggressor, while ColumnDisturb produces a weaker but extended population that remains above retention and changes abruptly at subarray boundaries. These observations motivate a spatial taxonomy based on aggressor–victim separation and array connectivity. Stress duration then constrains the physical interpretation: repeated short accesses continually reapply switching transients, whereas prolonged activation maintains the active electrical condition between transitions. The question is how the mechanisms available at each location act on the same storage-node state under these different access sequences.

Device studies provide several pieces of this picture. Trap-assisted electron migration can redistribute stored charge through carrier capture, emission and transport near an activated wordline, with strong sensitivity to trap position and energy [7]. Capacitive coupling perturbs the access-transistor electrostatics during wordline transitions [8,9], whereas sustained local electric fields can enhance transport during long activations [10]. Recent work has connected subsets of these mechanisms to RowHammer and RowPress observations [11,12], and circuit- and architecture-level models have incorporated activation count, row-open duration and leakage into vulnerability descriptions [13,14]. ColumnDisturb characterization, however, identified access-transistor and capacitor-to-bitline leakage as candidate mechanisms without resolving their device-level roles [4]. Connecting these local device processes to the extended failure distribution requires the bitline and sense-amplifier boundary conditions to be included alongside the aggressor wordline.

Here we combine dual in-line memory module (DIMM) characterization, device measurements and technology computer-aided design (TCAD) simulations to develop a unified device-level physical framework around three physical contributions to an evolving floating storage node. Trap-assisted electron migration describes local, defect-sensitive charge redistribution. Capacitive coupling describes transient perturbations of nearby wordlines and access-device electrostatics through parasitic capacitance and distributed wordline resistance. Sense-amplifier-mediated disturb describes the modification of unselected-cell leakage by the rail-level bitline bias imposed during sensing and restoration. Their different spatial ranges provide a physical interpretation of the measured spatial taxonomy: nearby victims can experience all three contributions, local EM and CC weaken with increasing separation, and SAMD can persist in remote cells connected through the active bitline/sense-amplifier network.

The same framework describes how the relative contributions evolve with stress duration. RowHammer repeatedly reintroduces wordline transitions and the active bitline condition. RowPress begins with a similar short-time response but extends the interval over which the aggressor wordline and bitline remain biased, allowing sustained leakage to accumulate after the initial transients decay. ColumnDisturb provides the complementary spatial limit, in which strong local wordline contributions are reduced while the driven-bitline condition remains. These temporal and spatial comparisons connect the different disturbance modes through their electrical boundary conditions and the resulting storage-node charge evolution.

Beyond this spatial taxonomy, data-dependent failure signatures provide further constraints on the proposed interpretation. These include the difference between first-failure onset and high-count RowHammer populations, the polarity observed during long-duration RowPress and ColumnDisturb, and the reversal in attack effectiveness between the neighboring wordline (NWL) and passing wordline (PWL) with increasing active time [4,11]. We examine these signatures through trap effectiveness, storage-state-dependent leakage and the changing balance between transient and sustained transport. They test whether the same mechanisms that explain the spatial distribution also account for the vulnerability and response of different cell populations.

The paper follows this progression. Section 2 describes the experimental and TCAD methods and establishes the measured spatial taxonomy. Section 3 interprets these distributions through the spatial ranges of EM, CC and SAMD, and explains how their relative contributions change between short RowHammer accesses and prolonged RowPress activations. Section 4 examines the additional population, polarity and aggressor-geometry signatures. Section 5 discusses the implications for reliability modelling, testable predictions and remaining quantitative questions, followed by the conclusions in Section 6.

## 2. Characterization methods and spatial read-disturbance taxonomy

We first describe the DIMM-level characterization and access-device measurements, then introduce the TCAD framework used to relate the imposed electrical conditions to storage-node evolution. The measured failure distributions subsequently establish the spatial read-disturbance taxonomy that constrains the physical interpretation in Section 3.

### 2.1. Experimental setup and measurement protocols

DIMM-level characterization is performed using a field-programmable gate array (FPGA)-based DRAM testing platform built on DRAM Bender [15], which provides low-level control over the DRAM command interface and enables programmable sequences of activate (ACT) and precharge (PRE) commands and fine-grained timing control. The platform allows the aggressor active time, precharge time, access count and data pattern to be independently controlled, so that RowHammer, RowPress, ColumnDisturb and retention can be evaluated within a common experimental framework. Before each disturbance test, the memory is initialized with the prescribed data pattern, the corresponding stress sequence is applied, and the victim region is subsequently read back to determine the number, location and polarity of the resulting bitflips. Unless otherwise specified, the DIMM-level measurements use a complementary data pattern, in which the victim rows are initialized to the opposite data value of the aggressor row. Figure 1a summarizes the DIMM measurement setup and DRAM hierarchy.

The DIMM-level measurements are performed at 85 $^\circ C$, with a 4-s retention interval used as the reference condition. RowHammer consists of repeated short ACT–PRE cycles with $t_{RAS} = 36$ ns and $t_{RP} = 14$ ns, for $8 \times 10^7$ activations. RowPress retains the ACT–PRE sequence but extends the aggressor ON time to 70.2 $\mu s$, followed by the same 14-ns precharge interval, for $5.68 \times 10^4$ activations. ColumnDisturb again uses $t_{RAS} = 36$ ns and $t_{RP} = 14$ ns for $8 \times 10^7$ accesses, but the activated row can be spatially remote from the affected victim cells while remaining connected through the shared column circuitry. These protocols deliberately vary both the temporal structure and spatial location of the applied stress while probing the integrity of charge stored in an unselected DRAM cell. Figure 1b summarizes the initialization, stress and readout sequence and the protocol settings for each mode.

The corresponding electrical boundary conditions are also important for interpreting the disturbance. Across the three active modes, the victim wordline (WL) remains OFF, while the storage node (SN) is initialized near either 0 or $V_{CORE}$ and then floats during the stress interval. Before activation, the bitline (BL) pair is precharged near $V_{CORE}/2$. After ACT, the sense amplifier (SA) resolves the bitline signal and drives the associated BL/BL-bar pair toward approximately 0 and $V_{CORE}$; during PRE, the bitlines return toward $V_{CORE}/2$. Which rail is experienced by a particular unselected cell depends on the stored data, the true/complementary-bitline mapping and the relevant BL/SA connectivity. Thus, even though the victim wordline remains OFF in all three disturbance modes, the floating storage node is subjected to substantially different wordline and bitline histories.

The access-device measurements complement the DIMM-level tests by characterizing the leakage response under controlled terminal biases. The retention-like and ColumnDisturb-like transfer characteristics are measured with a drain bias of 1.0 V and a body bias of −0.7 V, while the source bias is set to 0.5 V and 0 V, respectively. Sweeping the gate voltage probes the OFF-state response relevant to an unselected cell. These measurements test the effect of the bitline-side bias; the device-current comparison is presented with the SAMD analysis in Section 3.1.

**2.2. TCAD setup and calibration**

The TCAD representation couples the access-device geometry to the surrounding wordline and storage-node circuit, as summarized in Fig. 1(c,d,f). The local device domain includes the storage-node and bitline regions and the relevant neighboring or passing wordlines. Trap-containing cases represent carrier capture and emission in the local silicon; the external wordline network represents capacitive coupling and resistive relaxation. The sense-amplifier contribution is applied through the bitline voltage history. This separation connects the device, circuit and terminal-bias descriptions used in the mechanism analysis.

For device simulation, the following models were employed: Fermi–Dirac statistics, mobility models, high-field saturation models, and recombination models such as Shockley–Read–Hall (SRH) recombination with doping dependence, as well as band-to-band tunneling.

The DIMM-level electrical sequence defines the simulation boundary conditions. The victim wordline remains OFF, and the storage node is initialized near 0 or $V_{CORE}$ before floating under the imposed WL/BL history. The bitline is distinguished between the precharge level, $V_{CORE}/2$, and the active rail-level bias established by sensing. The evolving storage-node voltage, $V_{SN}$, provides the link between device transport and the DIMM-level disturbance signatures. Static bias snapshots characterize individual electrical conditions, whereas transient calculations follow the floating storage node through the applied sequence.

The representative wordline-coupling configuration uses 1.25-ns rise and fall times, $V_{PP} = 2.5\ V$, $V_{BBW} = -0.2\ V$, $R_{WL} = 1\ k\Omega$, $C_{SN} = 10\ fF$ and a nominal coupling parameter $C_C = 515\ fF$. Individual comparisons differ in aggressor selection, pulse duration, initial storage state and bitline bias. In particular, the representative 7.8-μs RowPress simulation hold is distinct from the 70.2-μs DIMM protocol in Section 2.1.

The baseline access device is calibrated against array-averaged leakage measurements from approximately 700,000 devices (Fig. 1e). The comparison focuses on the transfer characteristic around the nominal OFF-wordline bias, $V_{BBW} = -0.2\ V$, including the gate-bias range perturbed by wordline coupling. Matching the ON-state response alone would leave the leakage that governs storage-node evolution poorly constrained. This calibration establishes an average-device baseline; the role of rare, effective trap configurations is discussed separately in the RowHammer interpretation.

The calibration and the disturbance comparisons serve different purposes. The transfer characteristic constrains the baseline leakage response, while the spatial, temporal and polarity signatures in Sections 3 and 4 test the physical interpretation under changing boundary conditions. Agreement with an average transfer curve does not by itself establish the distribution of first-failure thresholds or validate every transient trap response.

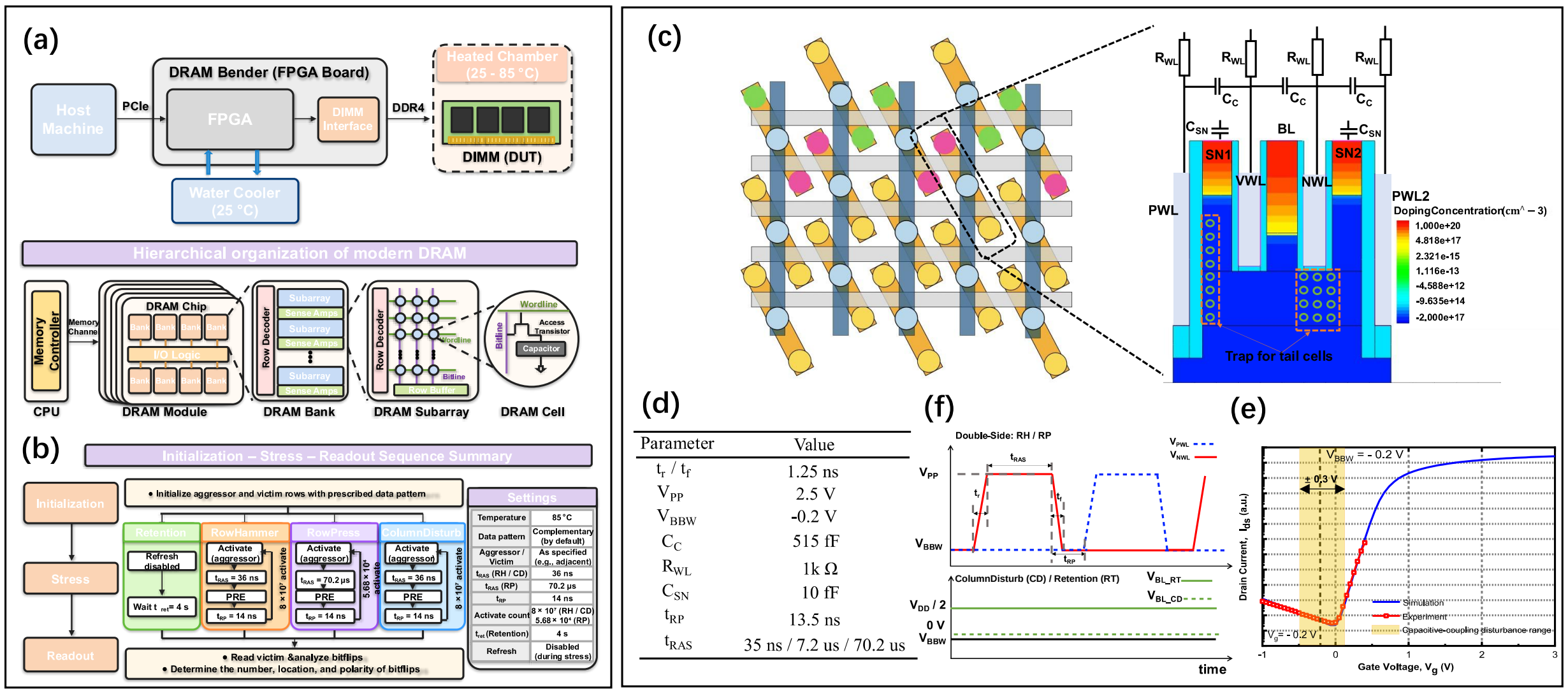

Figure 1. Experimental characterization and TCAD framework. (a) DRAM Bender setup for DIMM characterization and the hierarchical organization of DRAM. (b) Initialization–stress–readout sequence and protocol settings for RowHammer, RowPress, ColumnDisturb and retention. (c) Cell-array geometry, TCAD device representation, trap regions and equivalent wordline/storage-node network. (d) Device, circuit and timing parameters used in the simulations. (e) Measured and simulated access-device transfer characteristics, highlighting the gate-bias range around the OFF-wordline condition. Calibration uses array-averaged leakage from approximately 700,000 devices. (f) Wordline waveforms for double-sided RowHammer/RowPress and bitline biases for ColumnDisturb and retention.

### 2.3. Spatial read-disturbance taxonomy

With the measurement protocols and simulation framework defined, we first examine the DIMM-level spatial failure distributions. These observations establish the experimental constraints for the mechanism analysis that follows.

Figure 2 summarizes the spatial distribution of the failure population as a function of row address. The retention baseline remains close to zero across the measured region, whereas the three active disturbance modes exhibit clearly different spatial profiles.

RowHammer and RowPress produce highly localized failure peaks around the aggressor region in Subarray 3. The largest failure populations occur in the nearest victim rows, with the number of bitflips decreasing for victims located progressively farther from the aggressor. RowPress produces a similar localized spatial pattern, with strong peaks associated with the first-, second-and third-nearest neighboring rows. Thus, both RowHammer and RowPress are characterized by a pronounced near-aggressor failure population.

In contrast, ColumnDisturb exhibits a much broader spatial distribution. Rather than being confined to a small number of rows around the aggressor, failures extend across a large range of row addresses. The ColumnDisturb population forms approximately plateau-like regions within individual subarrays and changes abruptly at the subarray boundaries. The failure level is highest in the aggressor subarray and lower in the neighboring subarrays, while remaining clearly above the retention baseline over the electrically exposed region.

This spatial profile therefore defines three experimentally distinct regimes: a strong local population close to the aggressor, a weaker extended population within the surrounding array region, and a remote population extending into neighboring subarrays. The abrupt changes at the subarray boundaries further indicate that the remote failure distribution is not determined by physical row distance alone, but is associated with the underlying subarray and shared-column organization.

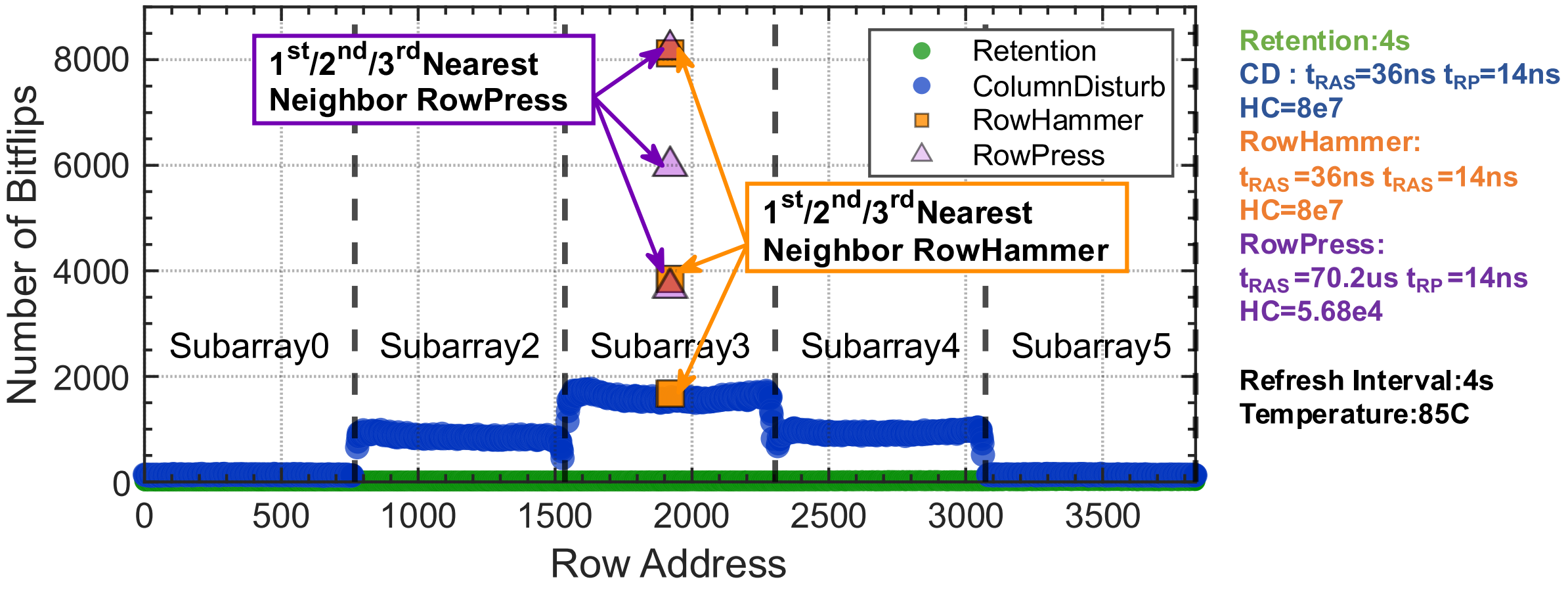

Figure 2. Spatial comparison of RowHammer, RowPress, ColumnDisturb, and retention failures at 85 °C. The ordinate gives bitflips per row, and the abscissa gives row address; dashed vertical lines mark the subarray boundaries as identified in the source data. RowHammer and ColumnDisturb use $t_{RAS}$ = 36 ns, $t_{RP}$ = 14 ns, and $8\times10^{7}$ activation cycles. RowPress uses $t_{RAS}$ = 70.2 μs, $t_{RP}$ = 14 ns, and $5.68\times10^{4}$ cycles. The nominal observation/refresh interval is 4 s. Local RowHammer/RowPress failures coexist with a broader ColumnDisturb population extending across the aggressor subarray and neighboring subarrays. The retention trace provides the same-duration baseline.

The spatial taxonomy in Fig. 2 provides the framework for the physical analysis that follows. Section 3 introduces the mechanisms responsible for these different spatial ranges. Electron migration, capacitive coupling and sense-amplifier-mediated disturb, illustrated in Fig. 3(a–c), Fig. 3d and Fig. 3(e–g), respectively, provide local, distance-dependent and shared-column contributions that can be mapped onto the failure populations identified in Fig. 2.

## 3. Physical mechanisms and interpretation of the spatial taxonomy

The DIMM-level characterization in Fig. 2 reveals a clear spatial hierarchy of read disturbance, ranging from strong failures close to the aggressor to weaker but persistent failures at much larger row distances and across subarray boundaries. To understand this spatial taxonomy, we propose three physical contributions to the evolution of the floating victim storage node: electron migration, capacitive coupling, and sense-amplifier-mediated disturb, illustrated in Fig. 3(a–g). These mechanisms originate from different parts of the DRAM device and circuit structure and therefore exhibit different spatial ranges. EM is associated with highly local field- and trap-assisted charge transfer; CC originates from parasitic coupling between wordlines and progressively weakens with their physical separation; and SAMD arises from the bitline voltage established by the sense amplifier and can therefore extend over much larger distances through shared column circuitry.

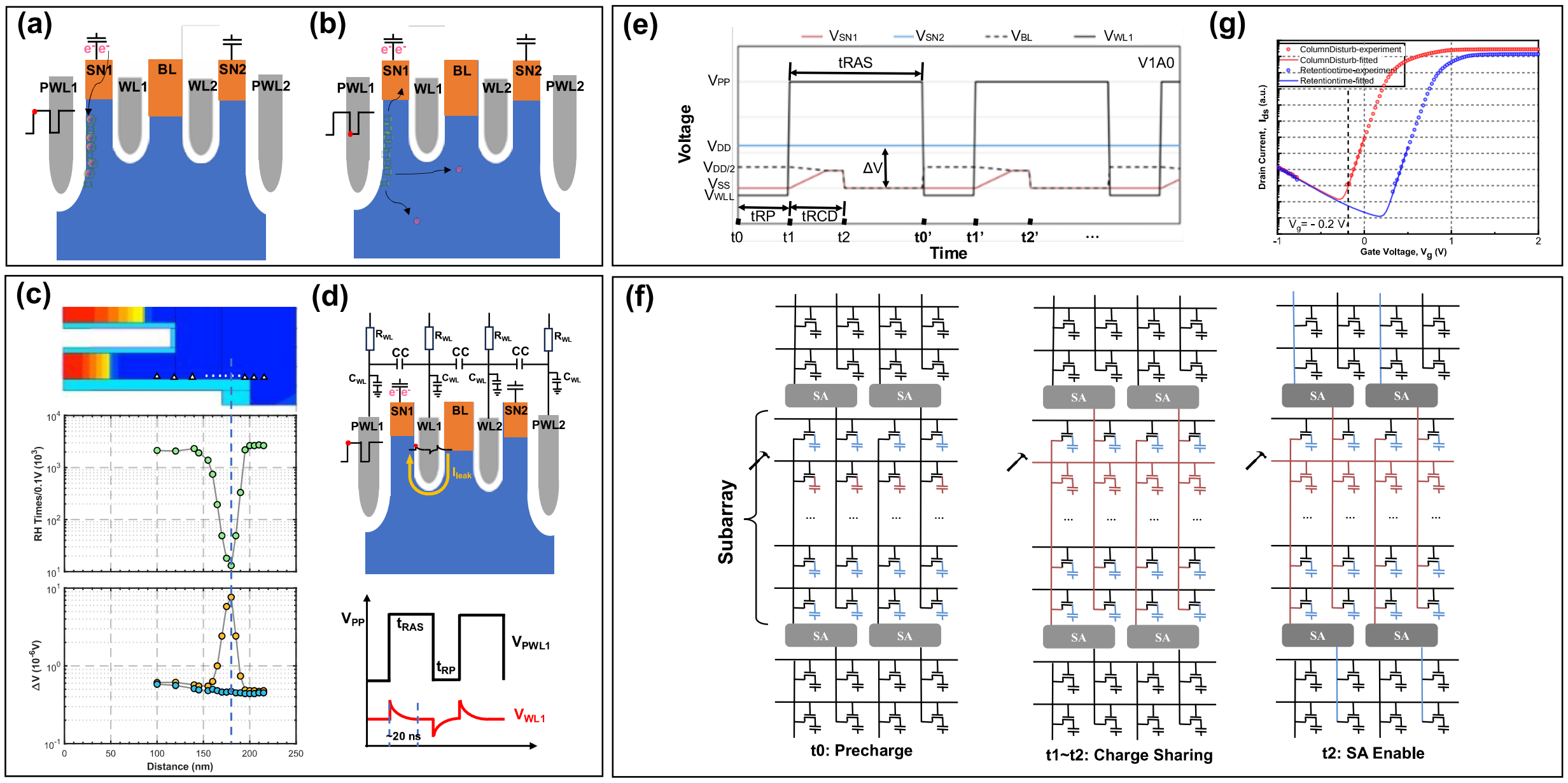

Figure 3. Physical mechanisms of DRAM read disturbance. (a) Trap-assisted electron capture. (b) Subsequent emission and redistribution. (c) Simulated trap-position dependence near the PWL: device geometry and swept trap sites, per-cycle storage-node voltage shift, and hammer cycles required for a 0.1-V shift under double-sided stress with a 36-ns aggressor ON time. Charge loss varies by orders of magnitude with trap position. (d) Capacitive coupling and the victim-wordline transient. (e) SAMD voltage sequence during row access. (f) Array connectivity during precharge, charge sharing and sense-amplifier enable. (g) Access-transistor characteristics under precharged (0.5 V) and SA-held (0 V) bitline biases. For a physical high storage state, the measured and fitted characteristics indicate approximately ninefold higher OFF-state current under the SA-held condition at $V_G = -0.2$ V. Schematics are not to scale.

### 3.1. Electron migration, capacitive coupling and sense-amplifier-mediated disturb

Electron migration denotes field- and trap-assisted carrier redistribution in the local silicon surrounding the aggressor and victim devices, as illustrated in Fig. 3(a,b). Activation of a neighboring or passing wordline modifies the local electric field and carrier concentrations, allowing defects near the affected region to capture carriers during the high-field phase and release them during subsequent voltage transitions. This capture–emission process can produce net charge exchange with the floating storage node [7,9].

Our simulations illustrate this positional sensitivity using traps near the PWL as an example (Fig. 3c). Depending on the trap position, the simulated charge loss can vary by orders of magnitude. Favorable positions produce larger per-cycle storage-node voltage shifts and require fewer hammer cycles to reach the same storage-node shift than less favorable positions. Trap presence alone therefore does not determine the strength of EM; its location within the local device environment is also critical.

This positional selectivity motivates the working hypothesis that rare, favorably positioned traps contribute to the highly vulnerable tail, whereas less effective configurations produce a weaker EM contribution. The corresponding tail-versus-bulk interpretation is developed in Section 4.1. Trap effectiveness can also vary with occupancy and capture/emission kinetics. Fluctuating trap occupancy has been proposed as a possible explanation for variable read disturbance (VRD) [16]; this temporal variability is beyond the scope of the present work.

EM is intrinsically local. The relevant trap-assisted charge redistribution requires the victim to reside within the local silicon region affected by the aggressor wordline field. Once the victim no longer shares this local device environment with the aggressor, the EM contribution disappears. EM can therefore contribute strongly to the nearest and most vulnerable cells but cannot account for the remote failures extending over large row distances in Fig. 2.

Capacitive coupling arises from the parasitic capacitance between an aggressor wordline and nearby victim wordlines, as illustrated in Fig. 3d. When the aggressor wordline switches during ACT or PRE, part of its voltage transition is capacitively coupled onto the victim wordline, producing a transient perturbation of the gate voltage of the victim access transistor. The amplitude and duration of this perturbation are determined by the distributed resistance and capacitance of the wordline network, and the perturbation gradually diminishes as the wordline driver restores the nominal OFF-state voltage through the resistance-capacitance (RC) network.

During this interval, the perturbed gate voltage modifies the electrostatics and leakage current of the victim access transistor, allowing additional charge exchange with the floating storage node. The resulting disturbance therefore depends both on the electrical properties of the wordline network and on the characteristics of the victim access transistor, including its threshold voltage, subthreshold response and leakage behavior. Importantly, the mutual capacitance between the aggressor and victim wordlines decreases as their physical separation increases. The coupled gate-voltage perturbation and the associated CC-induced disturbance therefore also decrease with increasing row distance. CC consequently extends beyond the immediately adjacent cell but progressively weakens as the victim moves away from the aggressor.

Sense-amplifier-mediated disturb is a circuit-level contribution often overlooked in device-level descriptions focused on local aggressor–victim interactions. As illustrated in Fig. 3(e,f), the bitline pair is precharged near $V_{CORE}/2$ before ACT. During sensing, the sense amplifier amplifies the selected cell's signal and drives BL and BL-bar toward opposite rails, approximately 0 and $V_{CORE}$. An unselected victim sharing this BL/SA path therefore experiences a rail-level bitline bias while its wordline remains OFF and its storage node floats.

This bias change modifies $V_{DS}$, $V_{GS}$, the source–body bias and the junction fields of the OFF access transistor, allowing additional charge exchange with the floating storage node throughout the active interval. For a physical high storage state, changing the bitline bias from $V_{CORE}/2$ (precharged) to 0 V (SA-held) increases the OFF-state current by approximately ninefold at $V_G = -0.2$ V, as indicated by our device measurements and fitted characteristics (Fig. 3g).

Some device-level analyses simplify this boundary condition. For example, Yang and Lin [7] held the bitline at a constant $V_{CORE}/2$ in their RowHammer simulations, approximating the precharge condition. This treatment omits the SA-driven rail-level bias during active access and can misrepresent the leakage of unselected cells. The role of this BL/SA-mediated contribution in nonlocal ColumnDisturb is discussed in the next subsection.

The three mechanisms have distinct spatial ranges. EM is confined to the aggressor's local silicon environment. CC propagates through wordline-to-wordline capacitance and weakens with increasing separation. SAMD follows BL/SA connectivity and can therefore affect both nearby and physically distant victims.

**3.2. Physical interpretation of the spatial taxonomy**

Having established the three physical contributions, we next use their different spatial ranges to interpret the failure-number-versus-row taxonomy in Fig. 2. We begin with ColumnDisturb because it provides the clearest experimental manifestation of the newly identified SAMD contribution, before considering the localized RowHammer and RowPress populations. Within these local populations, the access duration determines the relative accumulation of transient and sustained contributions.

ColumnDisturb. ColumnDisturb provides the remote limit of the spatial taxonomy. The affected cells can be located far from the activated row while their own wordlines remain OFF. At these distances, the victim does not share the local silicon region required for EM, and the parasitic coupling responsible for CC becomes negligible. The BL/SA connection, however, remains active. The remote victim can therefore continue to experience the SA-driven bitline condition and its associated leakage even after the local aggressor-related mechanisms have disappeared.

As shown in Section 3.1 and Fig. 3g, the SA-held bitline bias enhances OFF-state leakage relative to the precharged retention condition. This additional leakage explains why the remote ColumnDisturb population remains above retention in Fig. 2 and supports its interpretation as predominantly SAMD driven.

The subarray dependence provides further support for this interpretation. In the open-bitline architecture represented in Fig. 2, the sense-amplifier structures are shared by bitline segments associated with neighboring subarrays. The aggressor subarray is exposed to the full relevant set of active bitlines, whereas only a subset—approximately one half in the neighboring subarrays shown—shares the corresponding sense amplifiers. The exposed cell population therefore changes abruptly at subarray boundaries, producing the plateau-like spatial structure observed experimentally. The exact failure ratio also depends on cell vulnerability, but the stepwise spatial profile follows BL/SA connectivity rather than physical row distance, as expected for SAMD.

RowHammer. RowHammer represents the strongly local end of the spatial taxonomy. Victims immediately adjacent to the aggressor can experience all three contributions simultaneously. EM acts through highly local field- and trap-assisted charge transfer; CC perturbs the victim wordline through strong parasitic coupling; and SAMD modifies the OFF-state leakage through the active BL/SA condition. Their combined action provides a natural explanation for the much larger local RowHammer failure populations compared with remote ColumnDisturb.

Fig. 4a compares the modeled storage-node voltage change per cycle under double-sided RowHammer with EM + CC + SAMD, CC + SAMD and SAMD alone. Removing EM substantially reduces the voltage shift, while SAMD alone produces only a small response, indicating that EM dominates the simulated RowHammer response in the $6F^2$ saddle-fin structure studied here. This mechanism hierarchy may differ in $4F^2$ and 3D DRAM architectures, which we will address in separate work.

The relative contributions change systematically with row distance. EM disappears once the victim no longer shares the relevant local silicon environment with the aggressor. CC extends farther, but its magnitude decreases as the mutual capacitance between the aggressor and victim wordlines decreases. SAMD can persist through shared column circuitry. The decay of the RowHammer failure population with increasing row separation in Fig. 2 can therefore be understood as the progressive loss of the local EM and CC contributions on top of the more spatially extended SAMD background.

This spatial explanation does not, however, address why the earliest RowHammer failures preferentially occur in one polarity whereas the high-count population is dominated by the opposite polarity. That population-dependent behavior requires the additional role of trap effectiveness and storage-state-dependent leakage and is addressed in Section 4.1.

RowPress. RowPress is also strongly localized around the aggressor in Fig. 2, but its physical composition differs from RowHammer because the aggressor remains active for a much longer time. Victims close to the aggressor can again experience EM, CC and SAMD, accounting for the strong local failure population, but the relative weight of these contributions evolves during the long activation.

Fig. 4b compares two-cycle storage-node voltage trajectories for RowHammer and RowPress. During approximately the first tens of nanoseconds, the electrical condition is essentially that of a short RowHammer access. Wordline switching generates a CC transient, effective local traps can contribute EM, and the sense amplifier establishes the active bitline bias associated with SAMD. The initial RowPress response therefore contains EM + CC + SAMD.

As the aggressor remains active, however, the transient components diminish. The CC-induced victim-wordline perturbation decays as the wordline driver restores the OFF-state voltage through the distributed RC network, while the switching-related EM contribution becomes less important after the initial transition. In contrast, the sense amplifier continues to hold the shared bitline near a rail throughout the long active interval, while the aggressor wordline also remains biased. The storage node consequently continues to evolve through sustained leakage for the remainder of the hold. In Fig. 4c, the multi-microsecond interval produces a substantially larger accumulated storage-node shift than the initial ~35-ns transient, indicating that the total RowPress disturbance is dominated by the prolonged sustained-bias regime.

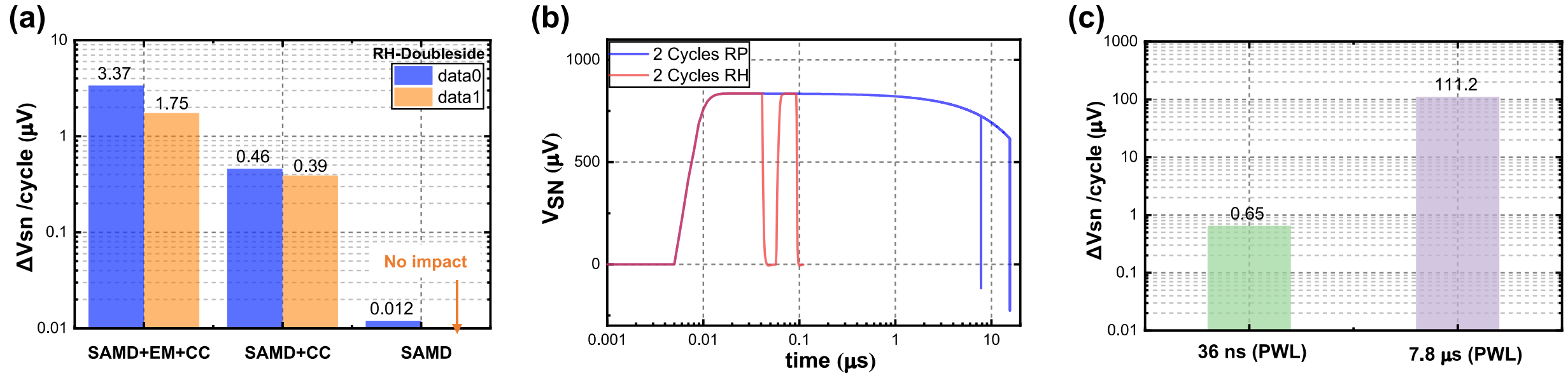

Figure 4. RowHammer decomposition and RowPress temporal evolution. (a) Storage-node voltage change per cycle under double-sided RowHammer in the modeled 6F² saddle-fin structure with EM + CC + SAMD, CC + SAMD and SAMD alone; the bars compare simulated configurations rather than independent additive contributions. (b) Storage-node voltage sequences over two cycles of RowHammer and RowPress. (c) Accumulated storage-node voltage shifts for a 36-ns PWL pulse and a 7.8-μs PWL hold, illustrating the transition from the initial transient toward sustained-bias leakage.

RowPress can therefore be viewed as a local disturbance that evolves from an initial RowHammer-like EM+CC+SAMD transient into a sustained-bias, leakage-dominated regime. This temporal evolution explains how RowPress can produce a strong localized failure population despite using far fewer ACT–PRE cycles than RowHammer. It also provides the physical basis for the RowPress polarity and aggressor-geometry signatures considered in Section 4.

Overall, the spatial taxonomy in Fig. 2 can be interpreted as a progressive change in the combination of the same three physical contributions. Immediately adjacent victims can experience EM + CC + SAMD; at larger row separations EM disappears and CC progressively weakens while SAMD remains; and at sufficiently large distances, including neighboring subarrays, the disturbance becomes predominantly SAMD driven. Within this spatial framework, the prolonged RowPress activation changes the relative accumulation of the available contributions, shifting the response from the initial switching transients toward sustained leakage.

Section 3 thus provides a physical interpretation of the spatial failure-number-versus-row taxonomy observed in Fig. 2. We next use the same device-level physical framework to address the remaining DIMM-level observations that cannot be explained by spatial range alone.

## 4. Explaining the remaining DIMM-level failure signatures

The spatial taxonomy establishes where EM, CC and SAMD can contribute, but several experimental signatures impose additional constraints on how these mechanisms evolve with cell vulnerability, stress duration and stored data. These include the opposite RowHammer trends at first-failure onset and high hammer counts, the disappearance of $0 \rightarrow 1$ failures during long-duration RowPress, the reversal of PWL/NWL attack effectiveness with ON time, and the data-dependent polarity of ColumnDisturb. We next examine each of these observations using the mechanism framework established in Section 3.

### 4.1. RowHammer: from the early vulnerable tail to the high-count population

Figures 5a and 5b reveal two apparently contradictory data-dependent RowHammer trends. Figure 5a shows the distributions of the hammer count to first failure, $HC_{\mathrm{first}}$: $0 \rightarrow 1$ bitflips emerge at lower hammer counts than $1 \rightarrow 0$ bitflips. In contrast, Fig. 5b shows that $1 \rightarrow 0$ failures outnumber $0 \rightarrow 1$ failures at high hammer counts. The first-failure threshold and the high-count population therefore probe different portions of the underlying cell-vulnerability distribution rather than representing two measurements of the same characteristic cell.

Guided by the trap-position dependence in Fig. 3c, we interpret this behavior using a tail-versus-bulk picture based on trap effectiveness. In this working hypothesis, highly effective trap configurations are rare within the overall population. The earliest failures therefore arise from an extreme vulnerable tail containing favorable trap configurations that can produce unusually large EM-driven charge transfer per activation. In the simulated double-sided RowHammer condition, PWL-side traps extract electrons from a physical low state more efficiently than the corresponding NWL-side traps inject electrons into a physical high state. The larger $0 \rightarrow 1$ per-cycle storage-node shift in the modeled tail configurations (Fig. 5c) is consistent with the earlier $0 \rightarrow 1$ onset observed in Fig. 5a.

As the hammer count increases, failures extend beyond this extreme tail into the much larger bulk-cell population. These cells do not contain comparably effective charge-transfer traps: traps may be absent, weaker, or less favorably positioned. Their weaker per-cycle response is represented by the bulk configurations in Fig. 5c, where CC and SAMD, together with the corresponding leakage response, govern the storage-node evolution.

The bulk results in Fig. 5c reveal an asymmetry in the dependence on storage-node state. As the data-0 storage-node voltage rises from 0 to $\Delta$ V, the per-cycle shift toward a $0 \rightarrow 1$ transition decreases substantially. The calculated current characteristics in Fig. 5d show how the changed storage-node bias modifies the leakage response, with the signed-current inset distinguishing charge-replenishment and charge-loss regions. The data-0 disturbance therefore becomes progressively self-limiting. In contrast, Fig. 5c shows little change in the data-1 per-cycle shift as its storage-node voltage falls from 1 V to $(1 - \Delta)$ V. This asymmetry supports the larger $1 \rightarrow 0$ failure population at high hammer counts in Fig. 5b within the proposed tail-versus-bulk interpretation.

The two RowHammer observations therefore arise from different regions of the vulnerability distribution. Rare and highly effective trap configurations determine the earliest $0 \rightarrow 1$ failures, whereas the broader CC/SAMD-driven population and its storage-state-dependent leakage response determine the high-count failure statistics. The observed polarity crossover does not require the same individual cell to change its failure mechanism as hammer count increases; rather, increasing stress progressively samples different portions of the cell population.

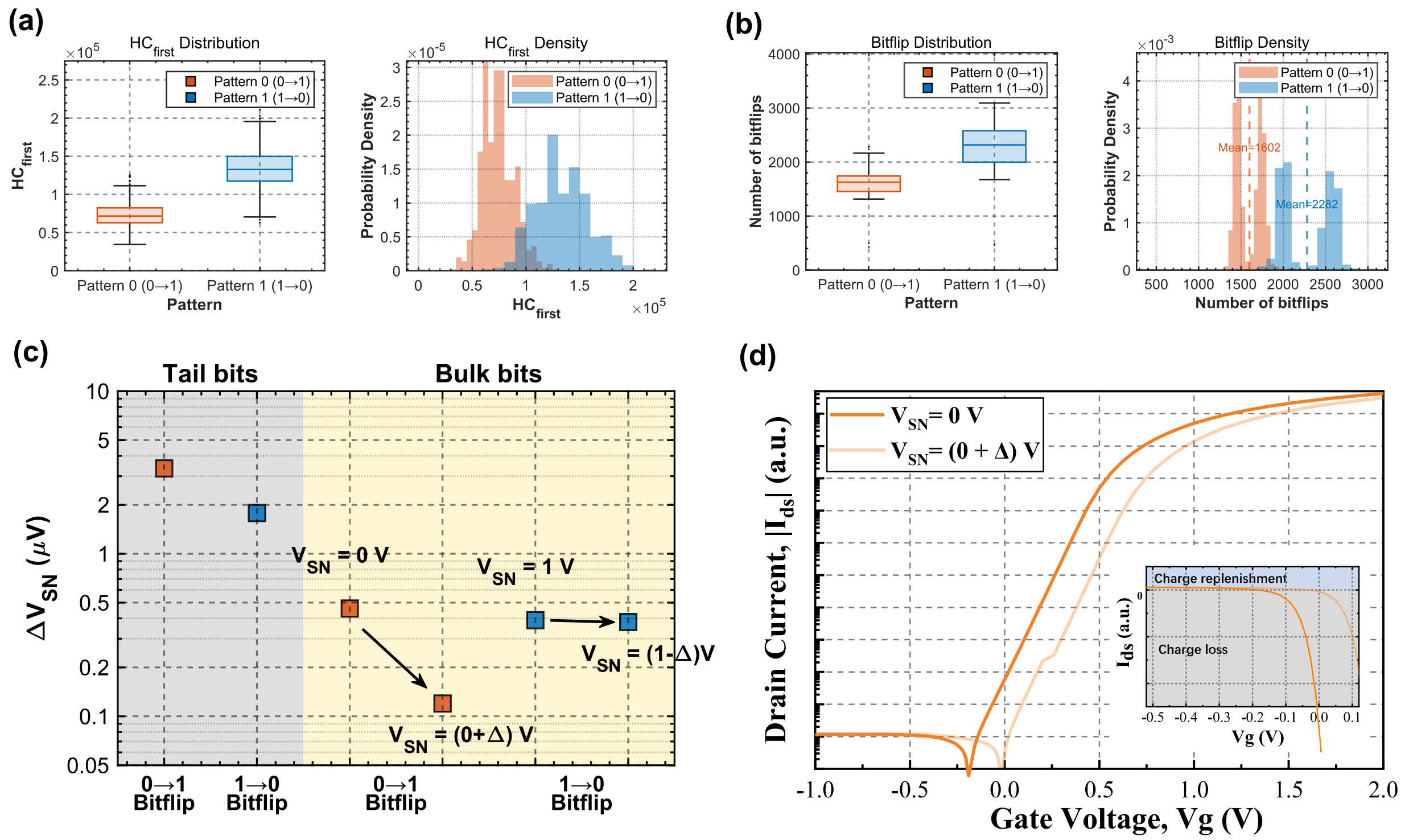

Figure 5. RowHammer first-failure onset and high-count failure populations. (a) Measured distributions of the hammer count to first failure, $HC_{\text{first}}$, showing lower thresholds for $0 \rightarrow 1$ than for $1 \rightarrow 0$ failures. (b) Measured bitflip-count distributions at high hammer counts, where $1 \rightarrow 0$ failures are more numerous. (c) Simulated per-cycle storage-node voltage shifts, $\Delta V_{\text{SN}}$, for representative tail and bulk configurations. The tail configurations show larger $0 \rightarrow 1$ than $1 \rightarrow 0$ shifts. In the bulk configurations, the $0 \rightarrow 1$ shift decreases as $V_{\text{SN}}$ rises from 0 to Δ V, whereas the $1 \rightarrow 0$ shift changes little as $V_{\text{SN}}$ falls from 1 V to $(1 - \Delta)$ V. (d) Calculated storage-node current magnitude versus wordline voltage for data 0 at $V_{\text{SN}} = 0$ and Δ V. The inset shows the signed-current regions corresponding to charge replenishment and charge loss, illustrating the bias dependence underlying the self-limiting data-0 response.

### 4.2. RowPress: consequences of the transition to sustained bias

The transition toward sustained-bias disturbance established in Section 3 and Fig. 4(b,c) is tested here against two RowPress signatures: the absence of $0 \rightarrow 1$ failures at long ON times and the reversal of PWL/NWL attack effectiveness. Figure 6a provides TCAD current maps for the polarity interpretation, while Fig. 6b compares the simulated aggressor ordering at short and long ON times.

#### 4.2.1. Why $0 \rightarrow 1$ failures disappear at long ON times

Under the tested long-duration RowPress conditions, $1 \rightarrow 0$ bitflips are observed whereas $0 \rightarrow 1$ bitflips are absent. The TCAD current maps in Fig. 6a provide a device-level interpretation: under sustained PWL and NWL biases, the indicated electron transport is consistent with accumulation at the victim storage node.

For a physical high-voltage state (upper row of Fig. 6a), electron accumulation reduces the positive storage-node charge and lowers $V_{SN}$, driving the cell toward a $1 \rightarrow 0$ failure. For a physical low-voltage state (lower row of Fig. 6a), the same tendency reinforces the low-$V_{SN}$ condition rather than moving the cell toward a $0 \rightarrow 1$ transition. The sustained leakage therefore acts asymmetrically on the two stored states: data 1 is destabilized, whereas data 0 is stabilized. This provides a physical explanation for the observed 1→0-only RowPress failures under the tested long-duration conditions.

The long-duration failure polarity therefore provides an independent test of the transition toward sustained-bias disturbance described in Section 3.

#### 4.2.2. Reversal of PWL/NWL attack effectiveness

A second RowPress signature is the change in the more effective aggressor geometry as the ON time increases. Luo et al. [11] report NWL-induced 1→0 failures under single-sided RowHammer and predominantly PWL-induced 1→0 failures under long-duration RowPress. Figure 6b shows a corresponding reversal in our simulated storage-node voltage shifts: NWL produces the larger shift for a 36-ns pulse, whereas PWL produces the larger shift for a 70.2-μs hold.

The short-pulse regime in Fig. 6b is RowHammer-like. For the 1→0 response considered here, NWL stress produces a larger storage-node voltage decrease than PWL stress, giving NWL > PWL under the modeled conditions. This is consistent with the NWL-induced 1→0 RowHammer failures reported by Luo et al. [11]. CC contributes through the wordline-switching transient, while EM can further enhance the response when an efficient trap configuration is present.

At long ON times, the initial switching transient represents only a small fraction of the total stress interval. The accumulated response is instead determined primarily by transport under the sustained aggressor and BL/SA biases. In the long-hold comparison in Fig. 6b, the held PWL configuration produces greater cumulative leakage in the data-1 cell than the corresponding NWL configuration, reversing the ordering to PWL > NWL.

The PWL/NWL crossover in Fig. 6b links the reported change in 1→0 failure behavior [11] to the evolving balance between switching-sensitive disturbance and sustained leakage. It supports the duration-dependent mechanism weighting described in Section 3.

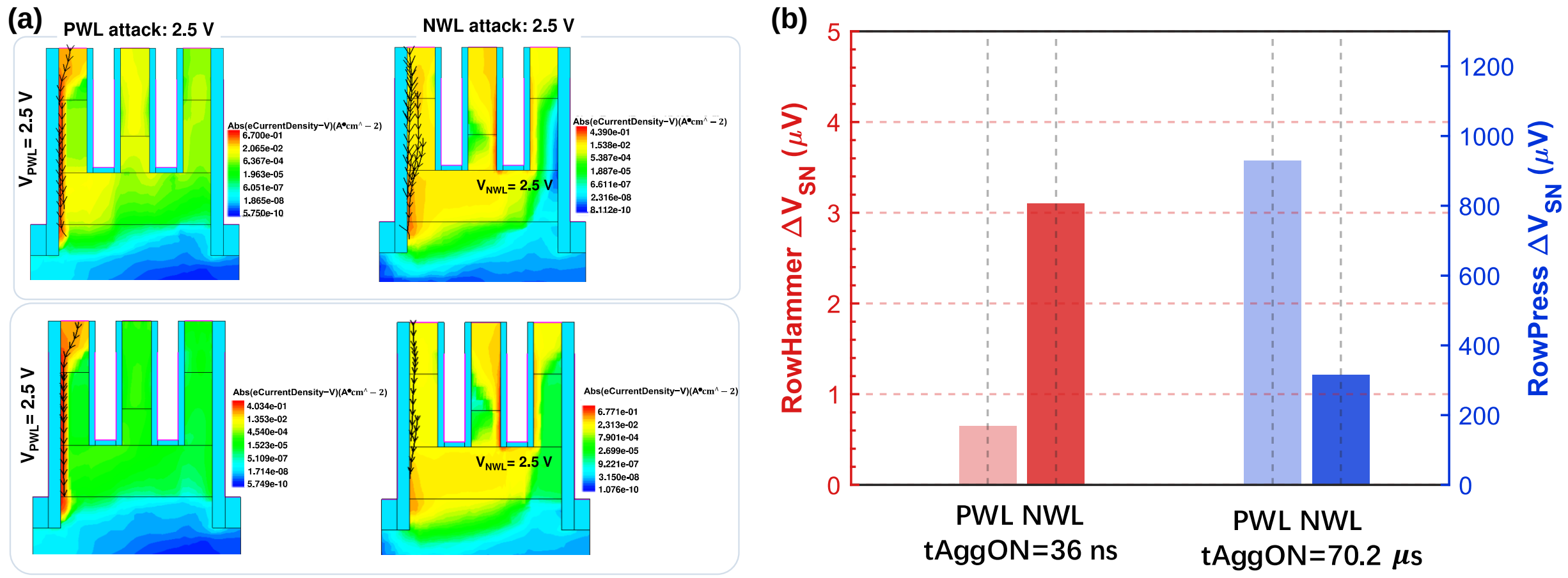

Figure 6. RowPress polarity and duration-dependent aggressor reversal. (a) TCAD electron-current-density maps and direction overlays for PWL (left column) and NWL (right column) biases of 2.5 V. The data-1 maps (top) show transport consistent with electron accumulation that reduces the positive storage-node charge and favors 1→0 failure; the data-0 maps (bottom) show transport that reinforces the low-voltage state and suppresses 0→1 failure. (b) Simulated storage-node voltage shifts under PWL and NWL attacks for 36-ns and 70.2-μs aggressor ON times, using the left and right vertical axes, respectively. NWL produces the larger short-pulse shift, whereas PWL produces the larger long-hold shift.

### 4.3. ColumnDisturb: data-dependent failure polarity

Remote ColumnDisturb is predominantly SAMD driven (Section 3). Here, we explain its data dependence: under the tested conditions, $1 \rightarrow 0$ bitflips are observed, whereas $0 \rightarrow 1$ bitflips are absent.

The data-0 TCAD condition in Fig. 7 provides a physical interpretation. With the victim wordline OFF, the storage node initially low and the shared bitline held high by the sense amplifier, the simulated current distribution and direction are consistent with a leakage response that favors electron accumulation in the low-voltage storage node. Rather than extracting electrons and driving the node upward toward a $0 \rightarrow 1$ transition, this response reinforces the low-$V_{SN}$ state. The $0 \rightarrow 1$ direction is therefore suppressed.

For the complementary physical high state, the corresponding low SA-held bitline produces an electrical condition that allows electron accumulation to reduce the positive storage-node charge, driving $V_{SN}$ downward toward a $1 \rightarrow 0$ failure. Thus, as in the sustained RowPress regime, the SA-driven boundary condition produces a data-dependent response in which the physical low state is stabilized while the high state is destabilized.

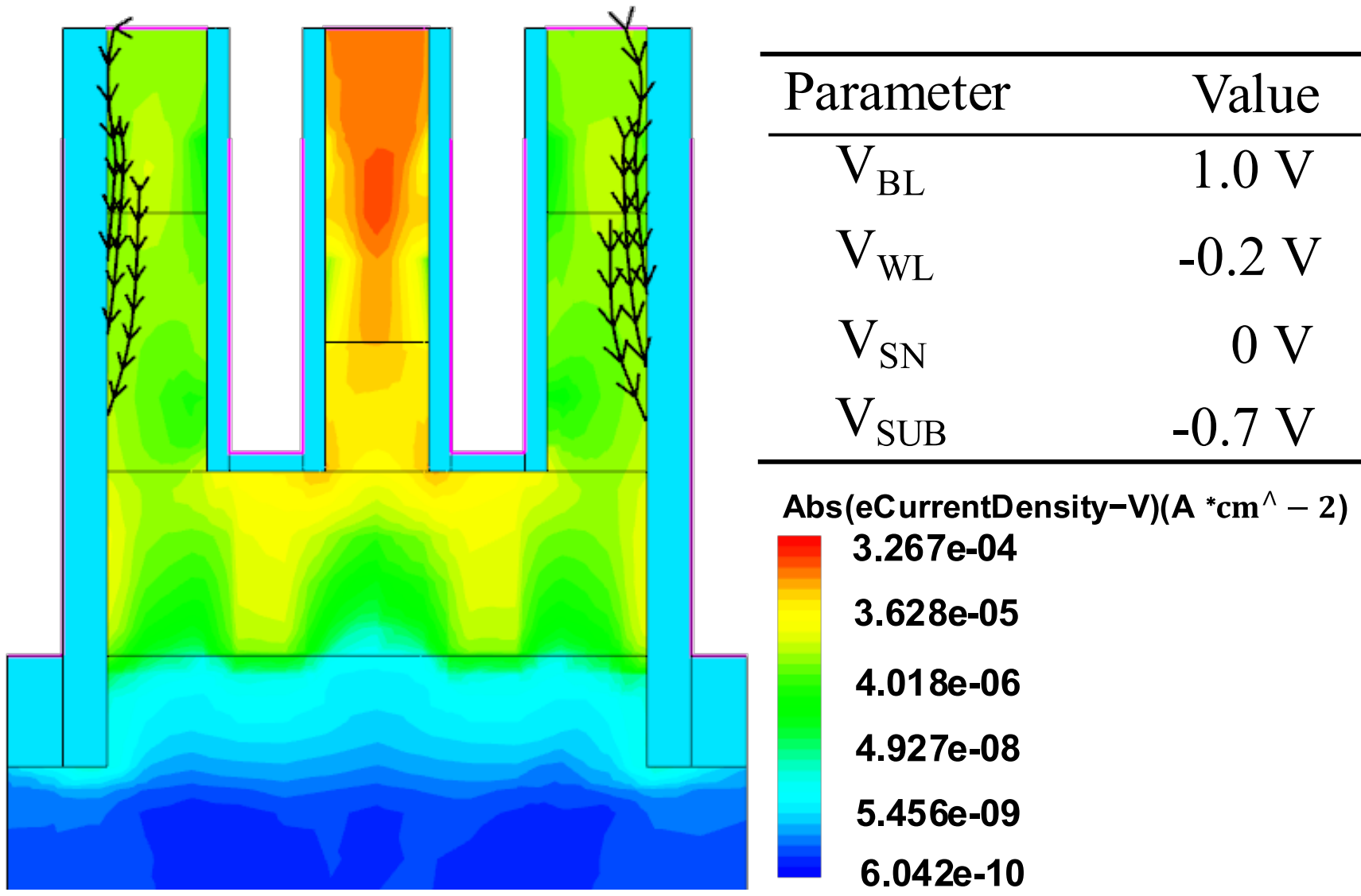

| Parameter | Value |
|---|---|
| $V_{BL}$ | 1.0 V |
| $V_{WL}$ | -0.2 V |
| $V_{SN}$ | 0 V |
| $V_{SUB}$ | -0.7 V |

Figure 7. Data-0 ColumnDisturb bias and polarity interpretation. TCAD uses VBL=1.0 V, VWL=−0.2 V, VSN=0 V and VSUB=−0.7 V with the victim wordline off. The electron-current-density snapshot and vector overlay support a leakage direction consistent with electron accumulation in the low-voltage storage node. Under the tested DIMM conditions, $1 \rightarrow 0$ failures are observed whereas $0 \rightarrow 1$ failures are not; the proposed SAMD response therefore stabilizes data 0 while destabilizing data 1.

The consistency between RowPress and ColumnDisturb is significant. In both cases, once the disturbance becomes dominated by sustained BL/SA-associated leakage, electron accumulation favors preservation of the low storage state and degradation of the high state. RowPress reaches this regime temporally by extending the active interval, whereas ColumnDisturb reaches it spatially by removing the local EM and CC contributions and isolating the shared-column response.

## 5. Implications for reliability modelling

The device-level physical framework provides a basis for translating device- and circuit-level mechanisms into analytical models for system-level reliability studies. Compact charge-transfer terms can describe storage-node evolution as a function of wordline and bitline history, stored state and array connectivity. Parameter distributions represent cell-to-cell variability, while a readout criterion links charge evolution to first-failure thresholds and bitflip populations.

Device measurements and TCAD can guide the form, sign and bias dependence of these terms, while DIMM measurements constrain their effective parameters and population distributions. The resulting compact description could evaluate long access sequences and assess mitigation strategies without repeated device simulation. Its predictive value should be tested on access patterns and operating conditions not used for calibration.

The analytical representation should preserve the physical dependencies relevant to the intended operating range. Developing and validating a complete predictive model is beyond the scope of the present work.

---

## 6. Conclusions

We have developed a unified device-level physical framework for DRAM read disturbance that connects DIMM-level failure signatures to device- and circuit-level charge-transfer processes. Three contributions—electron migration, capacitive coupling and sense-amplifier-mediated disturb—act on the same floating storage node but have different physical origins and spatial ranges. EM provides a highly local, trap-sensitive contribution; CC couples aggressor-wordline transitions into nearby victim wordlines and weakens with row separation; and SAMD arises from the rail-level bitline bias established by the sense amplifier and can therefore extend through shared column circuitry beyond the local aggressor environment.

Their spatial ranges explain the measured failure taxonomy, while their duration-dependent contributions and cell-to-cell variability explain the additional failure signatures. Adjacent RowHammer and RowPress victims experience the combined action of EM, CC and SAMD, whereas the remote ColumnDisturb population becomes predominantly SAMD driven and remains above retention because the active BL/SA condition enhances OFF-state access-device leakage. The measured $I_{CD}/I_{RT} \approx 9$ at $V_G = -0.2$ V provides direct device-level evidence for this circuit-induced contribution. With increasing RowPress ON time, the response evolves from an initial RowHammer-like transient toward sustained-bias leakage, explaining the long-duration polarity and the reversal of NWL/PWL attack effectiveness. In RowHammer, rare effective trap configurations determine the early $0 \rightarrow 1$ vulnerable tail, whereas the broader CC/SAMD-dominated population ultimately produces more $1 \rightarrow 0$ failures at high hammer counts.

Expressing these mechanisms as compact analytical charge-transfer terms offers a route from device-level physics to system-level reliability modelling. With calibration and independent validation, such models could support failure prediction and mitigation assessment across different access patterns and array configurations.